\documentclass[conference]{IEEEtran}
\usepackage{cite}
\usepackage{amsmath,amssymb,amsfonts}
\usepackage{algorithmic}
\usepackage{graphicx}
\usepackage{textcomp}
\usepackage{xcolor}
\usepackage[hyphens]{url}
\usepackage{physics}
\usepackage[font=sc]{caption}
\def\BibTeX{{\rm B\kern-.05em{\sc i\kern-.025em b}\kern-.08em
    T\kern-.1667em\lower.7ex\hbox{E}\kern-.125emX}}

\title{SnCQA: A hardware-efficient equivariant quantum convolutional circuit architecture}

\author{\IEEEauthorblockN{Han Zheng\IEEEauthorrefmark{1}\IEEEauthorrefmark{2}, Christopher Kang\IEEEauthorrefmark{1}, Gokul Subramanian Ravi\IEEEauthorrefmark{1}, Hanrui Wang\IEEEauthorrefmark{3}, Kanav Setia\IEEEauthorrefmark{2}, \\Frederic T. Chong\IEEEauthorrefmark{1}, Junyu Liu\IEEEauthorrefmark{2}\IEEEauthorrefmark{4}\IEEEauthorrefmark{5}\IEEEauthorrefmark{6}}
\IEEEauthorblockA{\IEEEauthorrefmark{1}Department of Computer Science, The University of Chicago, Chicago, IL 60637, USA}
\IEEEauthorblockA{\IEEEauthorrefmark{2}qBraid Co., Harper Court 5235, Chicago, IL 60615, USA}
\IEEEauthorblockA{\IEEEauthorrefmark{3}MIT EECS, Cambridge, MA 02139, USA}
\IEEEauthorblockA{\IEEEauthorrefmark{4}Pritzker School of Molecular Engineering, The University of Chicago, Chicago, IL 60637, USA}
\IEEEauthorblockA{\IEEEauthorrefmark{5}Chicago Quantum Exchange, Chicago, IL 60637, USA}
\IEEEauthorblockA{\IEEEauthorrefmark{6}Kadanoff Center for Theoretical Physics, The University of Chicago, Chicago, IL 60637, USA}
E-mail: \IEEEauthorrefmark{4}junyuliu@uchicago.edu}

\begin{document}
\maketitle
\thispagestyle{plain}
\pagestyle{plain}


\begin{abstract}

We propose SnCQA, a set of hardware-efficient variational circuits of equivariant quantum convolutional circuits respective to permutation symmetries and spatial lattice symmetries with the number of qubits $n$. By exploiting permutation symmetries of the system, such as lattice Hamiltonians common to many quantum many-body and quantum chemistry problems, Our quantum neural networks are suitable for solving machine learning problems where permutation symmetries are present, which could lead to significant savings of computational costs. Aside from its theoretical novelty, we find our simulations perform well in practical instances of learning ground states in quantum computational chemistry, where we could achieve comparable performances to traditional methods with few tens of parameters. Compared to other traditional variational quantum circuits, such as the pure hardware-efficient ansatz (pHEA), we show that SnCQA is more scalable, accurate, and noise resilient (with $20\times$ better performance on $3 \times 4$ square lattice and $200\% - 1000\%$ resource savings in various lattice sizes and key criterions such as the number of layers, parameters, and times to converge in our cases), suggesting a potentially favorable experiment on near-time quantum devices.

\end{abstract}

\section{Introduction}

\textbf{Quantum computing in the NISQ era: }Quantum computing is one of the most promising new computational technologies that could solve certain classically intractable problems \cite{shor1999polynomial}. In the current Noisy Intermediate-Scale Quantum (NISQ) era \cite{preskill2018quantum}, we are able to control hundreds to thousands of physical qubits, while quantum noises might limit the scale and capability of quantum circuits. On the other hand, potential near-term applications might still be useful regardless of noises before we arrive at large-scale, fault-tolerant quantum computation \cite{preskill1998fault} that is able to perform quantum error correction \cite{preskill1998reliable} faithfully.  

\textbf{Variational quantum algorithms: }Quantum machine learning \cite{biamonte2017quantum} enables quantum devices to run machine learning algorithms and obtain potential speedups, while variational quantum algorithms are leading paradigms to realize them in the NISQ era. Variational quantum algorithms \cite{cerezo2021variational} are made by shallow quantum circuits parametrized by trainable classical variational angles. Those variational angles can be optimized from some loss functions, similar to classical machine learning. Although suffering from quantum noises from devices and measurements, variational algorithms might be noise-resilient, and they have the potential to outperform their classical counterparts \cite{havlivcek2019supervised,liu2021rigorous}. Variational quantum algorithms are used in Variational Quantum Eigensolvers \cite{mcardle2020quantum} for quantum computational chemistry and quantum many-body physics, Quantum Alternating Approximation Algorithms (QAOA) \cite{farhi2014quantum} for quantum optimization, and quantum neural networks \cite{1} for quantum machine learning.  

\textbf{Symmetries in classical and quantum machine learning: }Symmetry is a ubiquitous phenomenon and a powerful tool to employ in physics and chemistry. Symmetry has emerged as a leading tool in the field of classical machine learning \cite{Cohen2016}. Incorporating symmetries into machine learning models by the name of equivariant neural networks, generalizing the concept of convolutional neural networks, has proven to be highly effective in many physical and chemical problems such as force fields learning of molecules \cite{CormorantNeurIPS} and learning ground states of quantum chemistry and quantum many-body models. Recently, there has been a surge of interest to borrow the insights from the classical equivariant neural networks to variational quantum algorithms by designing ans\"{a}tze respecting the underlying symmetry of the problems \cite{sauvage2022building, ragone2022representation, schatzki2022theoretical, Zheng2021SpeedingUL}. A key criterion for variational quantum circuits is if the given circuits could achieve \emph{computational quantum advantage} \cite{cotler2021revisiting}. Building symmetries into quantum variational ansatz, with appropriate state initialization such as performing quantum Schur transform \cite{krovi, harrow2005PhD, kirby}, is speculated to be immune to the so-called dequantization \cite{tang2022dequantizing}, which is stated in the recently-proposed permutational quantum computing plus (PQC+) model \cite{zheng2022super} in the case of global SU($2$) symmetry on qubits. The so-called SU($2$) symmetry on qubits concerns with the symmetry to applying rotations on every qubit or commuting to $U^{\otimes n}$ with $U$ is any single-qubit unitary. This is suggestive that equivariant quantum neural networks could lead to important insights about quantum advantage. 

\textbf{SnCQA: }In this work, we propose a unified class of variational quantum ans\"{a}tze that respects global SU($2$) symmetry and lattice automorphism-symmetry called the $S_n$ convolutional quantum alternating (SnCQA) ansatz with the number of qubits $n$ ($S_n$ permutation description is equivalent to SU($2$) in representation theory in this case). With theoretical foundations \cite{Zheng2021SpeedingUL,zheng2022super} set in hand, we add extra implementations to make it hardware-efficient and discuss practical perspectives on how SnCQA behaves as a computational architecture. First, we benchmark the performance of SnCQA, showing that it performs much better than the pure hardware-efficient ansatz (pHEA) \cite{kandala2017hardware,2}. Second, we benchmark the number of layers and parameters required to achieve fast training, showing that SnCQA requires much fewer resources compared to pHEA. Finally, through noise simulation, we show that SnCQA ansatz is scalable and resilient to quantum measurement noises. Our result indicates that incorporating symmetry into variational circuit architecture design could be a key principle in quantum machine learning, quantum chemistry, and physics problems of interest.  

\textbf{Key Results: }
\begin{itemize}
\item \textbf{Architecture design}. We design SnCQA architectures based on theoretical foundation established in \cite{Zheng2021SpeedingUL,zheng2022super} in the QAOA form, and design their hardware-efficient version. We show that the hardware-efficient versions are implementable in the near-term quantum devices. They are discussed in Section \ref{arch}. 
\item \textbf{Performance benchmarks}. We benchmark the performance of SnCQA compared to the traditional pure hardware-efficient ansatz (pHEA) on various homogeneous frustration-free antiferromagnetic Heisenberg model defind on 2D square lattices ranging sizes (row $\times$ column) from $2 \times 2, 2 \times 3, 2 \times 4$ and $3 \times 4$. We show that in our case, SnCQA could perform $20\times$ better than pHEA with the similar number of resources on $3 \times 4$ lattice, in terms of the precision to the ground state energy.  They are discussed in Section \ref{exp:perform}. 
\item \textbf{Resource benchmarks}. We benchmark the resource requirements of SnCQA compared to the traditional pure hardware-efficient ansatz (pHEA). We show that in our case, SnCQA could save from $200\%$ to $1000 \%$ hardware resources than pHEA with comparable performances. They are discussed in Section \ref{exp:resource}. 
\item \textbf{Noise resilience}. We benchmark the performance of SnCQA with shot noises simulated using IBM Qiskit \cite{ibmqiskit} and qBraid \cite{qbraid}. We find that SnCQA is sufficiently noise resilient in our cases even with significant amount of shot noises. They are discussed in Section \ref{exp:noise}. 
\item \textbf{Theoretical advantages}. We present theoretical arguments favoring SnCQA, including quantum neural tangent kernel \cite{Liu:2021wqr,liu2022analytic,liu2022laziness,wang2022symmetric}, universality \cite{Zheng2021SpeedingUL}, quantum advantage \cite{zheng2022super}, decoherence-free subspaces \cite{tran2021faster,tran2022digital,lidar2003decoherence} and saddle-point avoidance \cite{Liu:2022xdl}. They are discussed in Section \ref{theoretical}. 
\end{itemize}

%



\section{Background}

\subsection{Variational quantum algorithms}
\textbf{Brief introduction.} The variational quantum algorithms have emerged as the primary application in the NISQ area for solving various practical relevant questions such as combinatorial optimizations and challenging quantum chemistry and quantum physics problems. The key protocol to design good variational quantum algorithms is to build the quantum variational ansatz, which can be loosely described from two main categories: \emph{(a)} problem-independent ans\"atze and \emph{(b)} problem-specific ans\"atze. Problem-independent ans\"atze is normally characterized by its relative simplicity \cite{2, kandala2017hardware, barkoutsos2018quantum} and by its asymptotic expressibility over the Hilbert space $2^n \times 2^n$ \cite{biamonte2021universal, schuld2021effect}. The latter makes these ans\"atze useful for many problems while flexible in practical hardware-level implementation. For this reason, the problem independent ans\"atze are often referred to as the pure \emph{hardware efficient ans\"atze} or pHEA. The optimal expressibility, however, comes with the price of trainability, which fundamentally limits its practical effectiveness in many applications \cite{holmes2022connecting, cerezo2021cost}. On the other hand, there are problem-specific ans\"atze designs that take into account the specific structure of the problem, such as the quantum approximation optimization algorithms (QAOA) for solving combintorical puzzles. Taking inspiration from QAOA, a family of Hamiltonian variational ans\"atze has been proposed. These ansatzes are built typically according to the structure of the problem Hamiltonian and have been observed to mitigate trainability issues compared with pHEA \cite{wiersema2020exploring}. Quite naturally, one might take inspiration from the fundamental principle of nature in realizing a class of efficient ans\"atze, by exploiting the symmetry of the system of interests. Recently there have been works in building symmetry into variational quantum ans\"atze \cite{Zheng2021SpeedingUL, sauvage2022building, schatzki2022theoretical, nguyen2022theory}. In particular, in the case of SU($2$) symmetry, there have been a class of algorithms so-called equivariant quantum alternating ans\"atze built using the representation theory of the symmetric group $S_n$, which is speculated to possess super-exponential quantum advantage \cite{zheng2022super}. Quantum variational ans\"atze built from the principle of symmetry, effectively reduces the effective by even \emph{orders of magnitude} (see Figure~\ref{fig_scale} for instance of SU($2$) symmetry) while not harming the performance, suggesting a fruitful direction toward building effective and robust ans\"atze. 

\textbf{Noise.} We expect the presence of noise in the quantum circuits in the NISQ era; thus building ans\"atze that are resilient to the quantum noise are an essential consideration for any future large-scale implementation. There have been studies in the performance of the variational ans\"atze in presence of hardware noise \cite{wang2021noise, liu2022laziness}. In particular, noise could result in the barren plateu, severely limiting the trainability of the variational ans\"atze. Without full-scale employment of the fault-torelent quantum circuits, it is, perhaps, prudent to devise certain error mitigation techniques in implementing variational ans\"atze. By building symmetry-respecting ans\"atze with certain appropriate state initialization protocol, the effective dimension can be restricted to only the relevant subspace (also called charge sector) where the optimal solution would lie. This subspace, by the principle of symmetry, contains all necessary physical information of the problem, whose dimension could often be exponentially suppressed compared to the full Hilbert space (e.g., see Figure~\ref{fig_scale}) in tens of qubits, falling in the domain of the NISQ device. Furthermore, the system's underlying symmetry is known to be capable of encoding quantum information to these charge sectors in which to protect against various quantum noises, known as the decoherence-free subspaces \cite{tran2021faster,tran2022digital,lidar2003decoherence}. Therefore, it is natural to wonder the level of noise resilience the symmetry-respecting ans\"atze would possess by restricting the ans\"atze into the relevant charge sector of interest.

\subsection{Equivariant neural networks}

\textbf{Equivariance. }One of the most important neural network architectures in classical machine learning is Convolutional Neural Networks (CNNs) \cite{lecun1998gradient, krizhevsky2012imagenet,simonyan2014very,szegedy2015going,lecun2015deep}. In recent years, CNNs have also found applications in condensed matter physics and quantum computing. For instance, \cite{Lukin_2019} proposes a quantum convolutional neural network with $\log N$ parameters to solve topological symmetry-protected phases in quantum many-body systems, where $N$ is the system size. One of the key properties of classical CNNs is equivariance, which roughly states that if the input to the neural network is shifted, then its activations translate accordingly. There have been several attempts to introduce theoretically sound analogs of convolution and equivariance to quantum circuits, but they have generally been somewhat heuristic. The major difficulty is that the translation invariance of CNNs lacks a mathematically rigorous quantum counterpart due to the discrete spectrum of spin-based quantum circuits. For example, \cite{Lukin_2019} uses the quasi-local unitary operators to act vertically across all qubits. In quantum systems, there is a discrete set of translations corresponding to permuting the qudits as well as a continuous notion of translation corresponding to spatial rotations by elements of $\operatorname{SU}(d)$. Combining these two is the realm of so-called Permutational Quantum Computing (PQC) \cite{pqc}. Therefore, a natural starting point for realizing convolutional neural networks in quantum circuits is to look for {\it permutation equivariance}.

\begin{figure} [!ht] 
\includegraphics[width = \columnwidth]{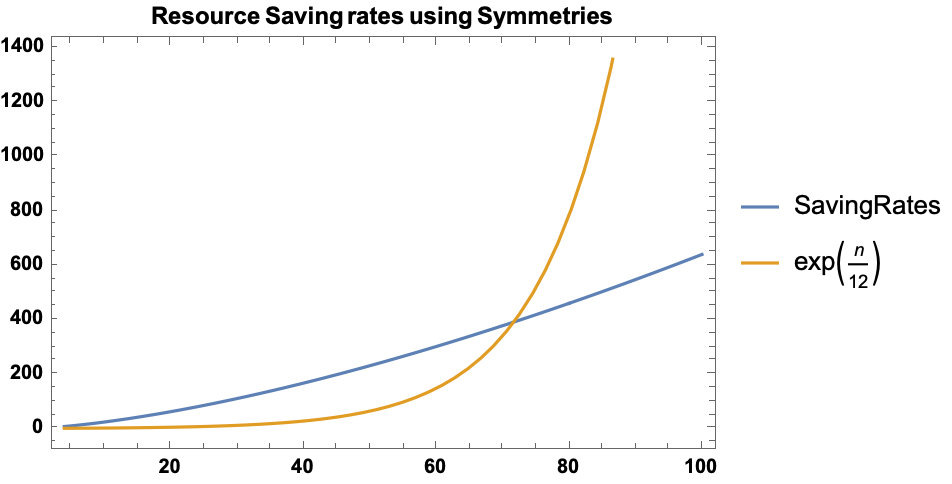} 
\centering
\caption{The scaling ratio between the Hilbert space dimension $2^n$ and the dimension of relevant charge sectors for ground states as $n$ goes from $4$ to $100$. Around 20-70 qubits, an exponential rate of saving in terms of dimensinality is observed.} \label{fig_scale}
\end{figure}

\textbf{Quantum neural networks. }Designing quantum neural networks aware of the problem symmetry (equivariant quantum neural networks) is an exciting emerging field, motivated by the success of classical equivariant neural networks. There are two major families of circuits: hardware-efficient as\"atze and problem-inspired ans\"atze. The former is designed to run on shallow depth consisting of one-and-two local gates, and the latter draws inspiration from quantum adiabatic computing, of which the quantum alternating approximation algorithm (QAOA) is an example of solving combinatorial problems. The family of ans\"atze that respects symmetries lies in between these two main families: such a family is only expected to work for problems themselves to exhibit the same type of symmetry, thus problem-specific in nature, but these ans\"atze might also be hardware efficient. We say that ans\"atz that is hard-efficient if it can be written as $\mathcal{O}(n)$ many at most 2-local gates. We define the symmetry group $G$ as the set of unitary operations that commute with Hamiltonian $H$, The explicit form of $G$ is problem-specific. In this paper, we consider the spatial lattice symmetry and global symmetry per interaction. The spatial lattice symmetry is derived from the automorphism group of the spatial interaction of Hamiltonian. The global symmetry is the additional symmetry that each interaction term of the Hamiltonian respects. In the case of the homogeneous Heisenberg Hamiltonian (see \cite{Zheng2021SpeedingUL,zheng2022super}), each interaction is SU($2$)-symmetric. Furthermore, the symmetry restricts Hamiltonian to the respective charge sectors in a block-diagonal manner by the Schur lemma. The dimension of the largest charge sectors has dimension $144$ on 12 sites possessing only roughly $ 4 \%$ of the Hilbert space dimension (See Figure \ref{fig_scale}). Thus it is intuitively useful to exploit the system's symmetry to design symmetry-protecting ans\"{a}tze that could lead to better approximations of the ground state and more robustness against experimental errors. Therefore, a key question to address is to derive a principled way to implement practical ans\"atz that could reinforce both kinds of symmetries yielding a good performance.

\section{Architecture design}\label{arch}

\subsection{SnCQA}

\begin{figure}[ht]
\centering
\includegraphics[width=\columnwidth]{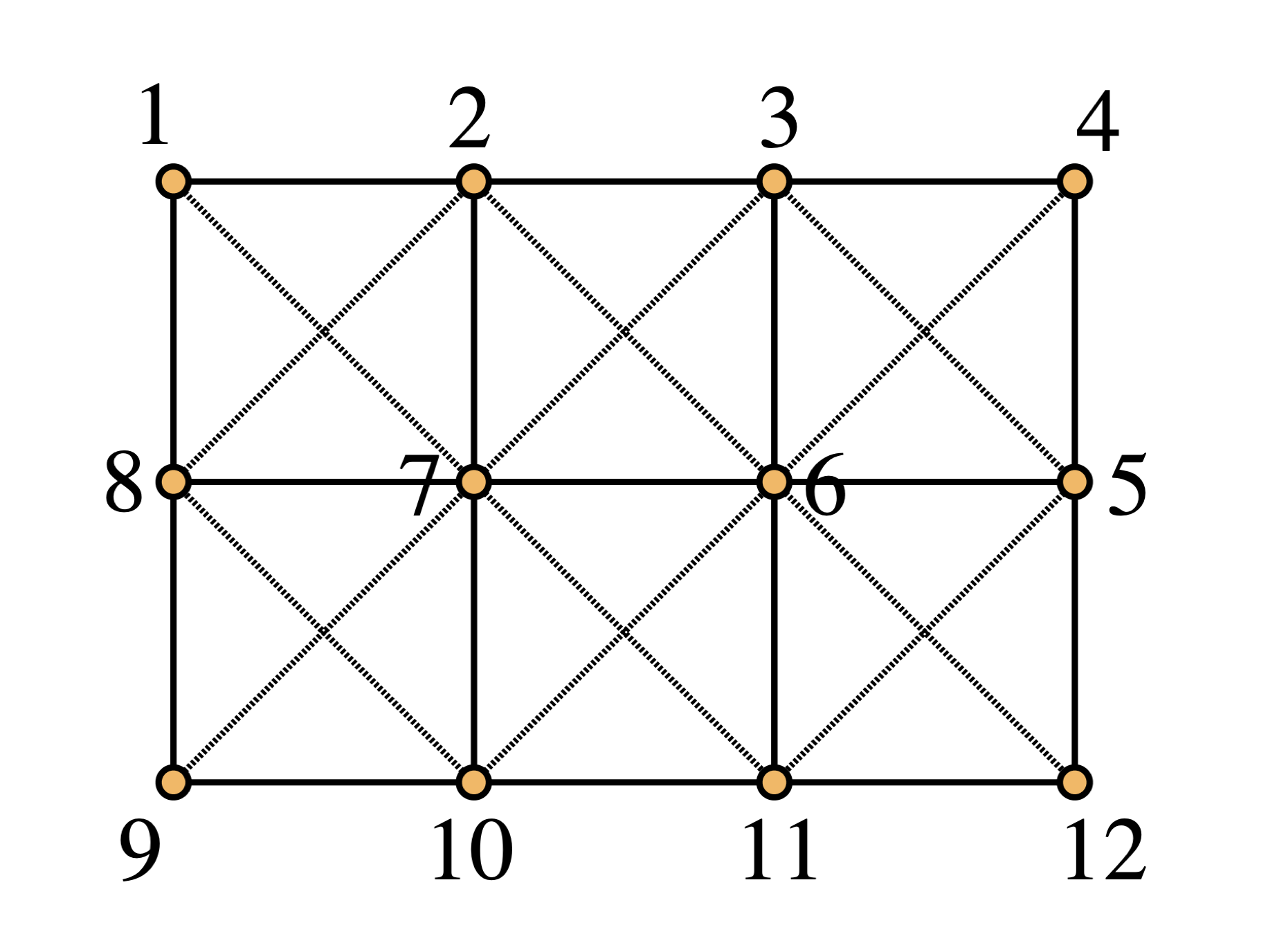}
\caption{$3 \times 4$ Square Lattice with nearest-neighbor and second-nearest neighbor pairing }
\label{Lattice}
\end{figure}

\begin{figure*}[ht]
	\centering
	\includegraphics[width=0.9\textwidth]{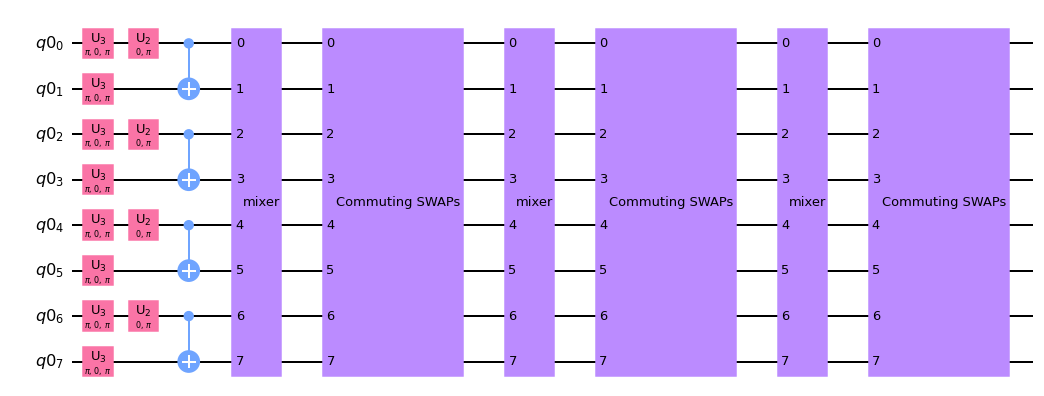}
	\caption{Complete Rountine for SnCQA with hardware efficient implementation}
	\label{fig: qiskitplot}
\end{figure*}

SnCQA takes inspiration from inspiration from the quantum approximate optimization algorithm (QAOA) with a \emph{driver Hamiltonian} that is the problem of interest and a \emph{mixer Hamiltonian} that is user-provided \cite{farhi2014quantum}. The QAOA ans\"atz takes $p$ many alternating layers of Hamiltonian simulation of the driver and mixer Hamiltonian. It can be shown that for sufficiently large $p$, QAOA ans\"atze recovers the quantum adiabatic computation. Thus, QAOA ans\"{a}tze is said to have guaranteed performance as the number of layers increases. Designing QAOA that respects symmetries has two major obstacles: \emph{(a)} how to construct symmetry-respecting Hamiltonians, and \emph{(b)} efficient protocol for state initialization. We address the above issues by proposing a novel class of QAOA circuits, drawing from the group theory, that could respect both the global SU($2$) symmetry of interactions and the spatial lattice symmetry. This becomes our paradigm of SnCQA.

The design of SnCQA comes from the Heisenberg $J_1$-$J_2$ Hamiltonian on the rectangular lattice (see \cite{Zheng2021SpeedingUL,zheng2022super}, and Figure \ref{Lattice}). In particular, we propose families of the SU($2$)-invariant Hamiltonians called $H_{\mathrm{YJM}}$, where each term is made by the \emph{Young-Jucy-Murphy} (YJM) elements \cite{Zheng2021SpeedingUL}. A particular advantage of $H_{\mathrm{YJM}}$ is that its ground state is very easy to be initialized, requiring only constant-depth circuits where we called $\ket{\psi_0}$. One can check that $\ket{\psi_0}$ projects to the charge sector with the same magnetization as the ground state of the unfrustrated regime where $J_2 = 0$. It is unknown that in the frustrated regime, which charge sector the ground state would lie. As a result, we generalize the constant-depth state initialization to each charge sector labeled by the total spin. We proposed the automorphism-invariant QAOA ansatz towards the $J_1$-$J_2$ Hamiltonian: $\{H_{i, 1}, H_{j, 2}\} $ where $H_{i, 1}$ takes the exchange elements of a given orbit $i$ of $Aut(H_1)$ the nearest-neighbor $J_1$ Hamiltonian and $H_{j,2}$ orbit $j$ the second-nearest neighbor $J_2$ Hamiltonian. The resulting ansatz, for which we call the SnCQA ansatz, is thus made by $H_{\operatorname{YJM}}$, $H_{i,1}$ and $H_{j,2}$. Note by the orbit-stabilizer theorem, the partition of $H_{i, 1}$ and $H_{j, 2}$ is disjoint, though the ansatz might not commute in general (see Figure \ref{cqa-architecture}).

\subsection{SnCQA: hardware-efficient implementations}
\begin{figure}[!ht]
\includegraphics[width = \columnwidth]{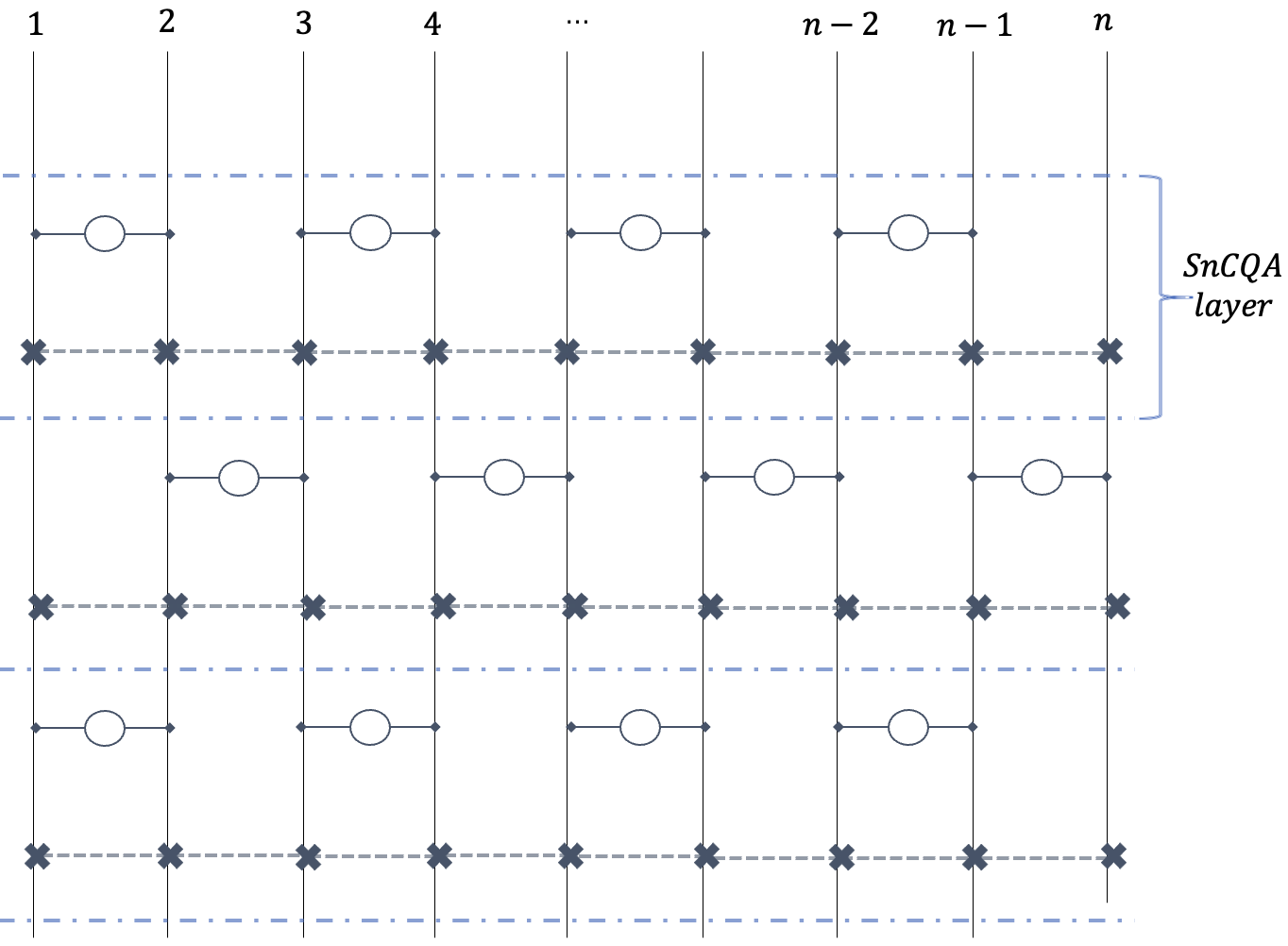} 
\centering
\caption{Schematic illustration of SnCQA with hardware-efficient implementation.} \label{cqa-architecture}
\end{figure}

The form of SnCQA discussed above has $\mathcal{O}(n^4)$ scaling circuit depths with the number of qubits $n$. Thus, we might consider a family of symmetry respecting ansatz that also could further decompose into geometrically more local SU($2$)-invariant gates in shallow circuits, where each layer would pack more independent parameters. In particular, let's consider the adjacent SWAPs $\tau_i$, which permutes the $i$-th and $(i+1)$-th qubit. We say $\tau_i $ is equivalent to $\tau_j$, denoted $\tau_i \sim \tau_j$ if there exists a unitary in the symmetry group that transforms $\tau_i$ to $\tau_j$. Then we consider the geometrically 2-local SU($2$) gates that are made by equivalence classes of $\tau_i$ with respect to the symmetry group. Combing with the $H_{\operatorname{YJM}}$, we obtain the hardware efficient implementation of SnCQA (See Figure \ref{cqa-architecture}). In our gate countings, we assume the standard primitive gates The dominant cost lies in implementing SnCQA is thus, in practice, depends on Trotter methods \cite{childs2021theory}. It can be shown in theory that the cost of the hardware-efficient SnCQA is of $\mathcal{O}(n^2)$ in \cite{Zheng2021SpeedingUL} using the linear combination of unitaries. 

\subsection{State Initialization}
State initialization becomes a key subroutine to perform. In particular, we should make sure that the state to be initialized to the span of the desired charge sectors. Common to \cite{Seki_2020, sauvage2022building, Zheng2021SpeedingUL}, the state initialization is proposed consisting of applying bit flipping $X$ followed by Hadmard and CNOT gates as shown in Figure~\ref{fig: qiskitplot}

\section{Experimental benchmarks}
\subsection{The model}
We consider the particular hardware-efficient ansatz design consisting of adjacent exponential (eSWAPs) and the so-called YJM mixer layer, depicted in Figure~\ref{cqa-architecture}. In particular, each layer consisted of the set of mutually commuting adjacent eSWAPs followed by the mixer layer given by the YJM Hamiltonian. In the practical simulation, we only consider the \emph{first-order} YJM Hamiltonian: $\sum^{n}_{i=2} \beta_i X_i$. It is shown \cite{zheng2022super} that the total circuit depth per layer scales as $\mathcal{O}(n^2)$. Furthermore, one could even cut off the number of terms in the YJM Hamiltonian by sampling a selection of YJM elements, which would improve the circuit depth by reducing the constant overhead.

\subsection{Performance benchmarks}\label{exp:perform}
We report the benchmark of performance for SnCQA with hardware efficient ansatz to that of ansatz proposed in \cite{kandala2017hardware, 2} on rectangular Heisenberg lattices of various sizes. The Hamiltonian consisted of the frustration-free antiferromagnetic Heisenberg Hamiltonian defined by nearest-neighbor interaction. It is shown analytically by the Marshall-Leib theorem that the ground state lies in the subspace given by the total magnetization $0$, which corresponds to the state initialization.

We observe that the SnCQA significantly outperforms the pure hardware efficient ansatz both in circuit depth and in training time needed to converge within $\epsilon$ precision to the ground state energy. The threshold value $\epsilon$ is set to be $0.05$ throughout the numerical comparisons.

\begin{table}[ht]
\centering
\caption{The number of layers and independent parameters needed for SnCQA to converge within $\epsilon$-precision to the exact ground state energies.  }
\resizebox{0.95\columnwidth}{!}{%
\begin{tabular}{|c||c c c c||} 
 \hline
 SnCQA &$2 \times 2$ & $2 \times 3$ & $2 \times 4$ & $3 \times 4$ \\ [0.5ex] 
 \hline\hline
 Layers &1 & 3 & 4 & 10$^\dagger$ \\ 
 Parameters \#&5 & 24 & 44 & 80$^\dagger$ \\ [1ex] 
 \hline
\end{tabular}
}
\label{table:cqa-layers}
\end{table}

The number of layers and parameters required by the SnCQA ansatz~\ref{cqa-architecture} to converge within $\epsilon$-precision to the exact ground state energies. The number of parameters is related to the number of layers by $(\frac{n}{2} + \gamma(n-1))p$, where $p$ is the number of layers, $n$ the number of qubits, and $\gamma \leq 1$ the sampling parameter of the YJM elements. We choose $\gamma =1$ throughout this section, except for in the case of $3 \times 4$, where we only randomly sample $3$ YJM elements out of a total of 11 different ones. Note that for $3 \times 4$ lattice, we used $4$ randomly sampled YJM elements which achieves within $0.7$-precision to the exact ground state energy. Though it would not match the convergence criterion, we note that it still significantly outperforms that of the pHEA method.

\begin{table}[ht]
\centering
\caption{The number of layers and independent parameters needed for pHEA to converge within $\epsilon$-precision to the exact ground state energies.}
\resizebox{0.95\columnwidth}{!}{%
\begin{tabular}{|c||c c c c||} 
 \hline
 pHEA &$2 \times 2$ & $2 \times 3$ & $2 \times 4$ & $3 \times 4$ \\ [0.5ex] 
 \hline\hline
 Layers &4 & 10 & 20 & 40$^*$ \\ 
 Parameters \#&48 & 180 & 480 & 1440$^*$ \\ [1ex] 
 \hline
\end{tabular}
}
\label{table:pHEA-layers}
\end{table}

The pHEA, pure hardware efficient ansatz, agnostic to the SU($2$) symmetry can pack more independent parameters per layer by a factor of $2$, with the number of parameter scales with the number of layers and qubits by $3np$. However, as the system size starts to grow, the performance of the pHEA drop significantly. In the case of $3 \times 4$ lattice, with $40$ layers and $1440$ independent parameters, the ansatz still could not reach within $\epsilon$-precision of the exact ground state energy, with the gap of roughly $13.4$. We fix the number of iterations to be $1000$ throughput while the learning rate is $0.1$. Another key criterion is the speed of the convergence: while running on the quantum device can be expansive in computational time, it is, therefore, very helpful to devise an ansatz that is constantly able to converge with few iterations as possible. We notice that, in this category, the SnCQA also significantly outperforms that pHEAs.

\begin{figure} [!ht] 
\includegraphics[width = \columnwidth]{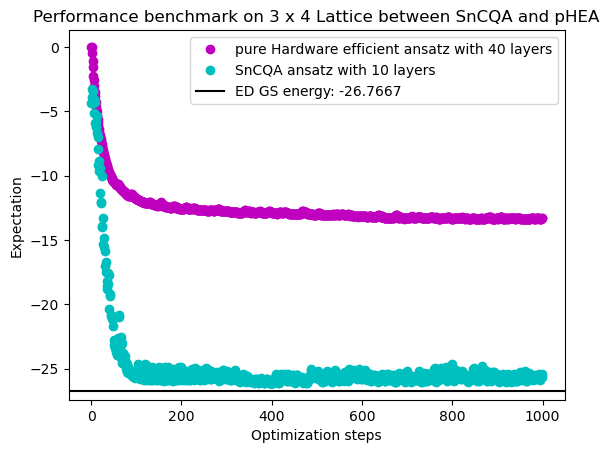} 
\centering
\caption{The Plot indicates that, while SnCQA under 10 layers with randomly sampled 3 YJM elements as mixer is unable to achieve $\epsilon=0.05$ precision (-26.102), its error is considerably smaller than that of the pHEA with 40 layers ($0.665$ vs. $13.383$) } \label{fig:3x4}
\end{figure}

\begin{table}[ht]
\centering
\caption{The minimal iterations needed for SnCQA and pHEA to converge within $\epsilon$-precision to the exact ground state energies. }
\begin{tabular}{|c||c c c c||} 
 \hline
  Time until convergence&$2 \times 2$ & $2 \times 3$ & $2 \times 4$ & $3 \times 4$ \\ [0.5ex] 
 \hline\hline
 SnCQA &16 & 27 & 65 & 360$^\dagger$ \\ 
 pHEA &33 & 55 & 219 & 1000$^*$ \\ [1ex] 
 \hline
\end{tabular}
\label{table:cqa-layers}
\end{table}
The number of layers used in SnCQA is $2, 4, 6, 10$ respectively and in pHEA $10, 20, 40, 40$ respectively. In the case of $3 \times 4$ lattice, the pHEA with $40$ layers could not converge even after 1000 iterations. We show the performance benchmark between SnCQA and pHEA in Figure~\ref{fig:3x4}, where $20\times$ separation of the relative precision to the exact ground state energy is observed.  $\dagger$ denotes that this is the observed time that SnCQA ansatz would estimate the ground state up to $\epsilon = 0.7$ precision.

\begin{figure} [!ht] 
\includegraphics[width=\columnwidth]{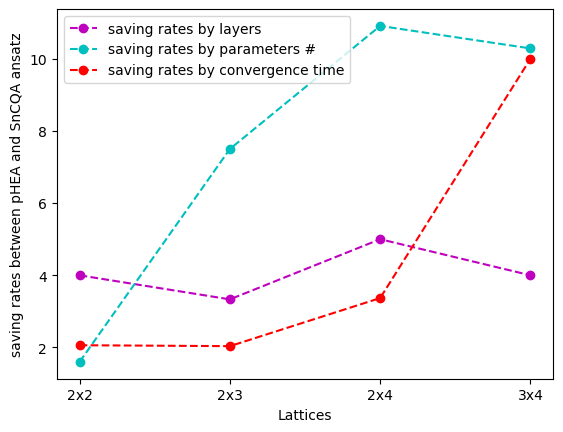} 
\centering
\caption{The rates between the key critical features (layers, parameters numbers, times) to converge to $\epsilon$ between the pHEA and SnCQA  ansatz studied in \cite{2, kandala2017hardware}.  } \label{fig: cqa_resource}
\end{figure}

A physically relevant and yet unsolved question is to look at the so-called $J_1$-$J_2$ antiferromagnetic Heisenberg model, while with $J_2/ J_1 \approx 0.5 $, there has been speculated novel phase called the spin-liquid phase. 

The spin-$1/2$ $J_1$-$J_2$ Heisenberg model consists of nearest-neighbor and second-nearest neighbor interaction defined on 2D lattices. The $J_1$-$J_2$ model has been the subject of intense research over its speculated novel spin-liquid phases at frustrated regions \cite{Balents2010}.  The unfrustrated regime $(J_2 = 0 \text{ or } J_1 = 0)$ for the anti-ferromagnetic Heisenberg model is characterized by the bipartite lattices, for which the sign structures of the respective ground states are analytically given by the  \textit{Marshall-Lieb-Mattis theorem} \cite{Lieb62}. The system is known to be highly frustrated when $J_1$ and $J_2$ are comparable $J_2/J_1 \approx 0.5$ \cite{Bukov2021} and near the region of two phase transitions from Neel ordering to the quantum paramagnetic phase and from quantum paramagnetic to colinear phase,  where no exact solution is known. Moreover, little is known about the intermediate quantum paramagnetic phase -- recent evidence of deconfined quantum criticality \cite{Nahum2015,Wen2016} sparked further interest in studying these regimes. Gaining physical insights in the intermediate quantum paramagnetic phase requires solving the problem of the ground state sign structure as the system approaches the phase transition. We report the performance SnCQA on the frustrated rectangular lattice with $J_2/J_1 =0.5$: 

\begin{table}[ht]
\centering
\caption{The number of layers, independent parameters, and iterations (time) needed for SnCQA to converge within $\epsilon$-precision to the exact ground state energies in the frustrated case $J_2/J_1 = 0.5$. }
\resizebox{0.8\columnwidth}{!}{%
\begin{tabular}{|c||c c c||} 
 \hline
 SnCQA &$2 \times 2$ & $2 \times 3$ & $2 \times 4$  \\ [0.5ex] 
 \hline\hline
 Layers &4 & 6 & 6  \\ 
 Parameters \#&12 & 42 & 48  \\ Times & 15 & 52  & 56 \\ 
 [1ex] 
 \hline
\end{tabular}
}
\label{table:cqa-layers}
\end{table}

Where the number of parameters is further reduced by randomly sampling $1$ YJM terms for $2 \times 2$ and $4$ YJM terms for $2 \times 3$ and $2 \times 4$ lattice size, while reaching the desired precision $\epsilon =0.05$. Throughout the entire section, all the algorithms are run with the same optimizer. 


\subsection{Resource benchmarks}\label{exp:resource}
Building an efficient ansatz that respects SU($2$) symmetry, however, comes with a price of possible worse scaling in terms of the circuit depth. Indeed, our SnCQA with hardware efficient ansatz scales as $\mathcal{O}(pn^2)$ Figure~\ref{fig: cqa_resource} where $p$ is the number of layers and $n$ the number of qubits. The gate counts are calculated by decomposing the circuits with the primitive gates: $\{\text{I}, \text{RZ}, \text{CNOT}, \text{X}, \text{H}, \text{RY}, \text{CY}, \text{CRY}, \text{SX}\}$, corresponding to the identity, controlled Pauli Z, CNOT, X, Hadamard, Pauli Y rotation, and controlled Pauli Y, controlled Pauli Y rotation, and the square root of Pauli X.  The eSWAP gates used here are implemented used in numerical simulation. In some implementations, the identity is removed as it only results in a trivial global phase factor. As a result, this implementation is also known as the exchange interactions \cite{bacon2000universal}.

The major cost lies in the implementation of the mixer layer by the first-order YJM Hamiltonian. As it is shown by some of us that the mixer YJM Hamiltonian serves as an analog to the mixer Hamiltonian in the QAOA \cite{Zhang_2021} setting where the eigenstate can be chosen as the so-called Schur basis by performing the efficient Schur transform \cite{krovi}. The design of the mixer Hamiltonian in QAOA is known to provide a theoretical guarantee as the number of layers increases. In our case, with varying lattice sizes, we observe that SnCQA results in a much faster convergence while packing much few parameters and layers. We note that the scaling could be made into sublinear and even linear time by only sampling fixed terms from the YJM Hamiltonian (i.e. set $\gamma = \mathcal{O}(1/{n})$). It would also be beneficial to consider the direct implementation by the so-called exchange interaction (eSWAPs with a difference in global phase) \cite{liang2018entangling}. Note that for the mixing layer by the YJM Hamiltonian, non-adjacent eSWAP needed to be implemented, though strictly 2-local. In our numerical simulation, the number of Trotter slices is $1$, which would likely result in Trotter errors that break the symmetry. However, we note that this error could be completely avoided if we could implement eSWAPs as one of the primitives, given its relatively simple form in the computational basis, where it can be further decomposed into a set of primitives without Trotterization by \cite{Seki_2020}.
Furthermore, we wish to comment that the Trotterization error can be even suppressed in our case by adapting the symmetry protection scheme considered in \cite{tran2021faster} by providing the unitary kicks given by $\{I^{\otimes n}, H^{\otimes n }\}$. 

\begin{figure} [!ht] 
\includegraphics[width = \columnwidth]{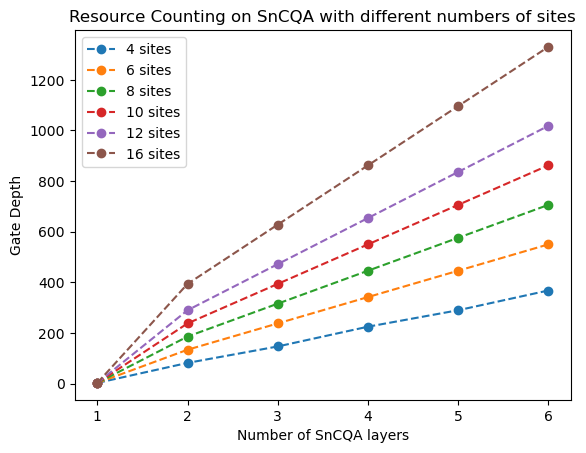} 
\centering
\caption{The circuit depth of SnCQA defined on Heisenberg lattices of varying number of sites (qubits). The plot indicates that the circuit depth scaling of $\mathcal{O}(pn^2)$. } \label{fig: cqa_resource}
\end{figure}

\subsection{Noise benchmarks}\label{exp:noise}
\begin{figure} [!ht] 
\includegraphics[width = \columnwidth]{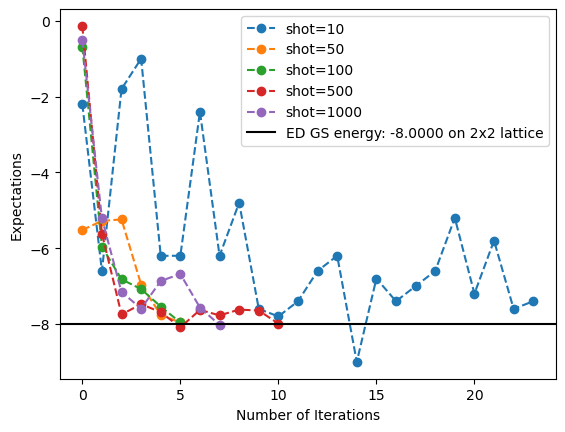} 
\centering
\caption{The plot indicates the resilience of SnCQA to the measurement noise with the number of shots ranging from $10, 50, 100, 500, 1000$. The higher the shots, the less noisy is in the measurement process.} \label{fig: cqa_noise}
\end{figure}

We also consider experimental noise in the simulation system. In particular, we consider the measurement noise for each time measuring the Heisenberg Hamiltonian during the gradient updates. The measurement noise is considered by the IBM state vector simulator: \UrlFont{$\operatorname{aer}{}_{}\operatorname{simulator} {}_{} \operatorname{statevector}$}, with the number of shots ranging from $10, 50, 100, 500, 1000$. The more shots the less noisy the measurement result would be. In Figure~\ref{fig: cqa_noise} we choose that even a circuit with noisy measurement would converge in very few iterations comparable to the noiseless simulation achieved by the pHEA ansatz. When the shots increase such as $50$, we note that the convergence time would be comparable to that of noiseless simulation by the SnCQA on $2 \times 2$ lattice, suggesting a degree of noise resilience. The implementation consisted of the hyperparameters: 2 layers, 1 trotter slice, randomly sample only 1 YJM element and learning rate of $0.1$.

\section{Theoretical advantages}\label{theoretical}
Aside from arguments of practical performances, theoretical studies suggest that equivariant quantum neural networks, including SnCQA, would have better performances \cite{Zheng2021SpeedingUL,zheng2022super,wang2022symmetric,sauvage2022building,ragone2022representation,schatzki2022theoretical,skolik2022equivariant,meyer2022exploiting}. Here, we propose some arguments supporting this claim for SnCQA.

\textbf{Quantum neural tangent kernel}. Quantum neural tangent kernels are theoretical tools to explore gradient descent dynamics for quantum neural networks \cite{Liu:2021wqr,liu2022analytic,liu2022laziness}. Higher quantum neural tangent kernel eigenvalues suggest faster convergences and possible better generalization performances with high alignments \cite{Liu:2022llk}. Thus, it is a good tool for analyzing variational quantum circuits with symmetries. In \cite{liu2022analytic,wang2022symmetric}, it is shown that quantum neural tangent kernel eigenvalues will generically scale linearly with the number of trainable parameters, and inversely with the square of the dimension of Hilbert space. For quantum circuits with symmetries, one can define effective Hilbert space dimensions \cite{liu2022analytic,wang2022symmetric} that reduce the dimension of Hilbert space towards the symmetric subspaces. It is consistent with the intuition that we can get better performances if we restrict the search space toward the symmetric subspaces. Similarly, it is shown in \cite{tran2021faster,tran2022digital} that respecting symmetry in Hamiltonian simulation by Trotterization leads to faster quantum simulation. This faster quantum simulation can be understood also by the reduction of effective dimension in simulation: dimensions of charge sectors are often polynomially less than the $2^n$ Hilbert space, if not exponential. 

\textbf{Quantum advantage}. 
Building on the framework of permutational quantum computing \cite{pqc}, it is proposed in \cite{zheng2022super} that there exists a super-exponential quantum speedup in evaluating matrix elements under the Schur basis with suitable locality assumption on its generating Hamiltonian. This quantum speedup is achieved by computing the so-called Fourier coefficients over the symmetric group, which is a classically difficult task with only factorial scaling proposed \cite{Clausen_1993}. It is, therefore, speculated in \cite{Zheng2021SpeedingUL} that the SnCQA would be useful in finding the ground states of the frustrated magnets in the NISQ between 20-70 qubits, where the exponential reduction of effective dimension is achieved in Figure~\ref{fig_scale}, for which classical algorithms such as the neural networks struggle to perform \cite{Westerhout2020}.

\textbf{Universality}. 
Furthermore, the universality result given arbitrary charge sector has been proved in \cite{Zheng2021SpeedingUL} using the adjacent eSWAPs and \emph{second}-order Hamiltonian. This offers great potential as it ensures the ansatz is expressible enough to find the ground state. While its effective dimension restricted by the symmetry might suggest a more robust answer to questions such as the barren plateau due to the suppression of effective dimension of the dynamical Lie algebra \cite{coles2021theory}. For the hardware efficient implementation of SnCQA, we note that the first-order YJM would not ensure the universality within a given charge sector, due to its optimal scaling of the circuit depth. Nevertheless, the numerical simulation with or without hardware noise suggests its Superior performance to that of pHEA by a great margin.

\textbf{Decoherence-free subspaces}. Charge sectors described by the system's underlying symmetry can be used to encode quantum information to the so-called decoherence-free subspaces \cite{lidar2003decoherence}, providing a potentially useful route to demonstrate robustness against experimental noise. Roughly speaking, symmetry itself provides protection against various quantum noises during the whole process of quantum machine learning (see, for instance, \cite{jiang2009preparation,friesen2017decoherence}). Thus, in our SnCQA construction, we expect that perhaps a similar mechanism could work, explaining the observed data with noise resilience. 

\textbf{Saddle-point avoidance}. Finally, it has been shown recently that quantum noises might be helpful in variational quantum algorithms due to the saddle-point avoidance mechanism \cite{Liu:2022xdl}. In classical machine learning, stochastic gradient descents are often performed, where noises are added manually to prevent possible trappings inside the regime of saddle points. Due to similar mechanisms, quantum noises are pointed out to be also helpful \cite{Liu:2022xdl}. In equivariant quantum machine learning, symmetries might lead to refined structures with lower numbers of saddle points, and those saddle points might be easier to get avoided due to symmetries.

\section{Summary}
Variational quantum algorithms in the NISQ era have the potential to outperform their classical counterparts and solve complex classical problems. In problems with intrinsic symmetries, variational quantum circuits will have shallower and more efficient designs. A generic type of symmetry is permutation symmetry, which widely appears in quantum chemical systems, graphs, and recommendation systems. In our work, we design SnCQA, a hardware-efficient variational quantum circuit architecture that respects $S_n$ permutation symmetries in quantum systems with $n$ qubits. We show that SnCQA could be implemented in a hardware-efficient way, and they are scalable, accurate, and noise-resilient in certain quantum systems. Using an example of the Heisenberg model in quantum mechanics, we benchmark the performance of SnCQA about their convergence, resource estimation, and noise resilience. We show that SnCQA could perform $20\times$ better, with $200\% - 1000\%$ resource savings, compared to pHEA, the traditional hardware-efficient quantum circuits without respecting symmetries. Moreover, our work shows that SnCQA is noise-resilient. 

SnCQA is a promising design for future realizations of NISQ algorithms in problems with symmetry. Along the line of our research, broader applications could be implemented in quantum computational chemistry, quantum optimization, and quantum machine learning in real problems \cite{Li:2022unf}.

\section*{Acknowledgment}
We thank Yi Ding, Jens Eisert, Laura Gagliardi, Liang Jiang, Risi Kondor, Zimu Li, Zihan Pengmei and Sergii Strelchuk for helpful discussions. This work is funded in part by EPiQC, an NSF Expedition in Computing, under award CCF-1730449; in part by STAQ under award NSF Phy-1818914; in part by NSF award 2110860; in part by the US Department of Energy Office of Advanced Scientific Computing Research, Accelerated Research for Quantum Computing Program; and in part by the NSF Quantum Leap Challenge Institute for Hybrid Quantum Architectures and Networks (NSF Award 2016136) and in part based upon work supported by the U.S. Department of Energy, Office of Science, National Quantum Information Science Research Centers. FTC is Chief Scientist for Quantum Software at ColdQuanta and an advisor to Quantum Circuits, Inc.
JL is supported in part by International Business Machines (IBM) Quantum through the Chicago Quantum Exchange, and the Pritzker School of Molecular Engineering at the University of Chicago through AFOSR MURI (FA9550-21-1-0209). This research also used resources of \texttt{qBraid.com} \cite{qbraid}.



\bibliographystyle{utphys}
\bibliography{refs}

\end{document}